# Autoencoding Coordinate Sequences from Psychophysiologic Signals

T. Hutcheson[1] and A. Raj[1]

1. Florida Institute for Human and Machine Cognition, 40 S Alcaniz St, Pensacola, FL 32502, USA
thutcheson@ihmc.org

*Abstract*— We present a method for converting 24 channels of psychophysiologic time series data collected from individual participants via electroencephalogram (EEG), electrocardiogram (ECG), electrodermal activity (EDA), respiration rate (RR) into trackable three dimensional (3D) coordinates sufficient to estimate participation in specific task and cognitive states.

## I. Introduction

Currently, identification of human cognitive state requires the collection of data from indirect measures such as electroencephalogram (EEG), electrocardiogram (ECG), electrodermal activity (EDA), respiration rate (RR), and other psychophysiological signals that represent the individual's autonomic nervous system balance. Once acquired and pre-processed, the underlying cognitive state signals can be observed and interpreted by human experts or more recently, by using machine learning (ML). Extracting cognitive state from the ML processed data allows for automated processing and tracking of cognitive state [1]. Accurate, real time cognitive state detection and tracking would allow for dynamic manipulation of parameters to maximize task performance or for evaluation of the efficacy of task-based interventions and countermeasures for inducement of intended cognitive state changes.

## II. Methods

### A. Overview

The present method attampts to track the cognitive state of a subject from electromagnetic (EM) energy received by the EEG transducer coils. We take advantage of the ability of an autoencoder (AE) to compress multidimensional data and, in the process, learn useful embeddings or approximations from the data [2]. As part of a larger study, we synchronously collected (300 samples/sec) EEG, ECG, EDA and RR using an integrated data acquisition system (DSI-24, Wearable Sensing, LLC, San Diego CA) along with task performance scores during a baseline three-minute resting-state (eyes closed) task and a controlled visual working memory task set (N-back) to create exemplars of high and low workload and stress cognitive states [3, 4].

This research is based upon work supported in part by the Office of the Director of National Intelligence (ODNI), Intelligence Advanced Research Projects Activity (IARPA), via N66001-24-C-4505. The views and conclusions contained herein are those of the authors and should not be interpreted as necessarily representing the official policies, either expressed or implied, of ODNI, IARPA, or the U.S. Government. The U.S. Government is authorized to reproduce and distribute reprints for governmental purposes notwithstanding any copyright annotation therein.

### B. Design

For this initial implementation, we curated a training data set consisting of an ensemble of data collected from a random subset of participants (n=3) during the eyes closed task.

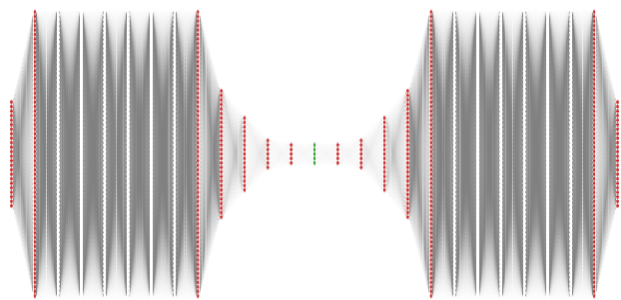

Fig. 1. Multi layer AE with sequential layers consisting of nodes (red) as follows: 128, 128, 29, 17, 7, 5, latent layer (green), 5, 7, 17, 29, 128, 128. The 128x128 node layers are illustrated in a folded format to accommodate for the large difference in layer dimensions. A 5-node buffer layer was inserted on each side of the latent layer to isolate the output.

We then constructed an autoencoding algorithm to orthonormalize at each training epoch and extended the mean standard error (MSE) of reconstruction cost function with an SVD-based angular function that drives the included angle of each of the eigenvectors to within +/- 0.3 degree of mutual orthogonality. By replacing the first epoch of training data with a repeating pattern of common mode EEG sensor data (i.e., channels A1 and A2), we established the signal to noise ratio (SNR) floor that adusted the weights of the AE prior to processing the time series data.

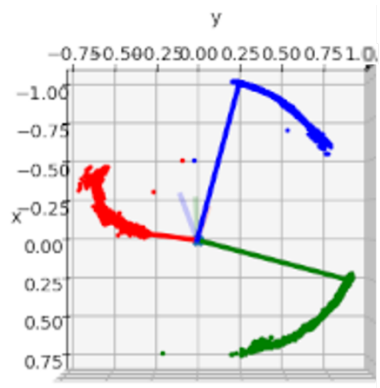

Fig. 2. Rotation of basis viewed along the + z-axis (red). The AE training concludes when all three vectors achieve orthogonal within +/- 0.15 deg.

Simultaneously the AE rotates the blue (second greatest magnitude) and green (smallest magnitude) traces are rotating into 90 degrees. The AE was constructed in PyTorch (MetaAI, New York, NY) using linear layers and rectifying logical unit (RelLU) activation at the larger 128 node layers to enhance the response to the inherently non-linear data. The remaining layers narrow steeply to a 5-dimensional latent space. This narrowing provides the essential high compression rate, creating a "Markov blanket" that loosely surrounds the embedding, preventing overfitting the bulk layer, instead [5][6]. Once trained, the AE maps the DSI-24's 21-channels of EEG, ECG, EDA and RR data collected during the workload and stress tasks into estimates of position displacement with respect to the manifold trained on the pooled eyes closed. The relative displacement from this "resting state" manifold should differ spatially by task condition (e.g., high or low for stress and for workload).

*C. Neuroanatomical basis*

Neural communication "hubs," represent areas of the brain that manage high numbers of links between brain regions and are known to consume a large fraction of the energy of the brain [7]. These sites would be expected to emit much more EM energy under cognitive load. For a given task, these hubs could influence the pattern of EM energy captured by the EEG biased toward the centroid of these hubs.

## III. RESULTS

*A. Preliminary Results*

In order to assess individual cognitive state displacements from the resting state manifold, we processed the psychophysiological data from each task through the trained AE and applied a sliding, 100 msec median filter over the output for the task. When plotted (see Fig 3) the traces demonstrate spatially different distribution for the high (H) and low (L) tasks.

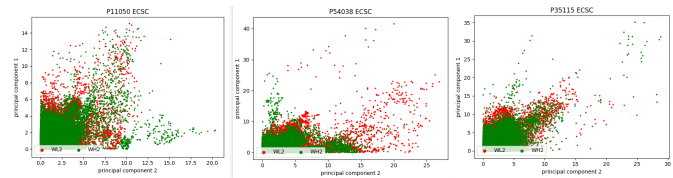

Fig. 3. AE output for three participants from high and low workload (WH1 & WL1, respectively) psychophysiologic data. The WH1 (green) manifests a distribution more compact than WL1 (red).

Interestingly, at this resolution, multiple participants manifest similar activation patterns. In these plots we see a distinct pattern in the AE representation of the data, but some subjects performing the same task, show a different pattern. The distance between clumps of activation are noted to correspond approximately to the layout and distances between the neural commination hubs, as described by [7]. To further explore the feature space, we extended the median filter to six seconds to reduce clutter and transients for visual observation of the acceleration and velocity of the coordinate sequence over time. This increases the spatial separation between the high and low workload tasks, providing insight into avenues for investigation for development of real-time detection and classification of cognitive state (see Fig 4).

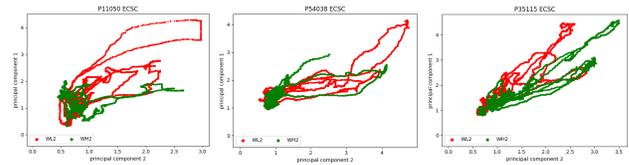

Fig. 4. Same participant data as Fig 3, but plotted with 6 sec median filter kernel illustrating increased separation and trackable trajectories.

## IV. CONCLUSION

This novel AE-based approach facilitates real time identification and tracking of cognitive state. Training the baseline manifold to a small subset of the total cohort provides sufficient information for the AE to converge, while reducing the volume of training data needed from each individual. This can increase generalizability and activity-specific classification with respect to training on a larger (non-cohort) population of individuals and reduces brittleness due to over-training only to individual data. Further development will seek to optimize the AE to tune to individual data, identify a cognitive state classification metric, characterize method accuracy, improve processing speed, and explore mapping the latent space to biophysical spatial constraints and neural communication hubs.

ACKNOWLEDGMENT

The authors thank Dr. Toshiya Miyatsu for his assistance and insights during development of this approach.